\documentclass[conference]{IEEEtran}
\IEEEoverridecommandlockouts

\usepackage[font=small]{caption}
\usepackage{cite}
\usepackage{amsmath,amssymb,amsfonts}
\usepackage{algorithm}
\usepackage[noend]{algpseudocode}  
\usepackage[bookmarks=false]{hyperref}
\PassOptionsToPackage{bookmarks=false}{hyperref}

\usepackage{graphicx,enumitem}
\usepackage{textcomp}
\usepackage{xcolor}
\usepackage{mathcomSTEv4}
\usepackage{url}
\usepackage{hyperref}

\usepackage{pgfplots}
\usepackage{pgfplotstable}
\pgfplotsset{compat=newest}
\usepackage[labelfont=normalfont,textfont=normalfont]{caption}
\newcommand{\algo}{\texttt{PolarZero}}
\newcommand{\Csf}{\mathsf{C}}

\def\BibTeX{{\rm B\kern-.05em{\sc i\kern-.025em b}\kern-.08em
    T\kern-.1667em\lower.7ex\hbox{E}\kern-.125emX}}
\begin{document}
\title{
Reinforcement Learning-Aided Design of \\ Efficient Polarization Kernels
\thanks{Y. Hong and S. Rini are with the Department of Electrical and Computer Engineering, National Yang Ming Chiao Tung University (NYCU), Hsinchu, Taiwan (emails: \{eltonhong.ee12, stefano.rini\}@nycu.edu.tw). 
L. Barletta is with the Department of Electronics, Information and Bioengineering (DEIB), Politecnico di Milano (PoliMI), Milan, Italy (email: luca.barletta@polimi.it).}
}

\author{
Yi-Ting Hong, Stefano Rini, and Luca Barletta
}

\maketitle
\begin{abstract}
Polar codes with large kernels achieve optimal error exponents but are difficult to construct when low decoding complexity is also required. 
We address this challenge under recursive maximum likelihood decoding (RMLD) using a reinforcement learning approach based on the Gumbel AlphaZero algorithm. 
The resulting method, \algo, consistently matches exhaustive search in identifying low-complexity kernels, and discovers a size-16 kernel with complexity comparable to handcrafted designs. 
Our results suggest that \algo{} is a scalable tool for large-kernel design, where brute-force search is no longer feasible. %
\end{abstract}


\section{Introduction}

Polar codes, introduced by Arıkan \cite{Ari09}, achieve capacity for any binary-input symmetric memoryless channel. The original construction uses a \(2 \times 2\) kernel, but larger kernels can achieve improved error exponents.
However, designing large kernels that simultaneously offer good error exponents and low decoding complexity becomes increasingly challenging due to the exponential growth of the search space. In this work, we propose a reinforcement learning (RL) approach to address this challenge.
Specifically, we focus on the design of kernels that (i) approach the optimal error exponent and (ii) minimize decoding complexity under recursive maximum likelihood decoding (RMLD). 
To this end, we implement a version of the Gumbel AlphaZero algorithm that efficiently explores the space of polarization kernels.
We refer to the resulting algorithm as \algo.
We show that, for small kernel sizes, \algo \space recovers known handcrafted low-complexity designs. 
This suggests that it can scale to larger kernels where brute-force methods become impractical. 

\noindent
\emph{\textbf{Related Work.}}
Two main aspects of polar code performance are relevant to this work: (i) error exponents and (ii) decoding complexity.

\noindent
\textbf{(i)} Regarding error exponents, \cite{KorErrExp09} studies large-kernel polar codes and shows that any \( \ell \times \ell \) matrix without an upper triangular column permutation polarizes symmetric channels. The authors derive bounds on the achievable error exponent and show that no kernel of size \( <15 \) exceeds the original \( 2 \times 2 \) kernel's exponent of \( 1/2 \). A BCH-based construction is proposed that asymptotically approaches exponent 1. 
In \cite{pdp_bound}, code decompositions are used to construct non-linear kernels with improved exponents. They provide optimal constructions for sizes 14–16 by meeting a new upper bound. 
Finite-length scaling behavior is studied in \cite{HasScaExp2x214}, where the blocklength required to maintain a target error rate as rate approaches capacity is shown to scale as \( (I(W)-R)^{-\mu} \), with \( \mu \approx 3.579 \). 

\noindent
\textbf{(ii)} From a complexity perspective, recursive maximum likelihood decoding (RMLD) is based on trellis-based ML decoding, originally introduced in \cite{FujRMLD98}, and builds on dynamic programming frameworks such as \cite{ForneyMITnotes}. The first application to polar codes appears in \cite{TriTre19}. 
The recursive trellis processing algorithm (RTPA) in \cite{TriRTPA23} generalizes this approach for large kernels by computing log-likelihood ratios through a recursive trellis structure, achieving lower complexity than full Viterbi decoding. In \cite{TroBrute24}, a depth-first search strategy improves lower bounds on the exponent for kernel sizes 17–29, producing kernels compatible with RTPA. 
Handcrafted kernels optimized for efficient decoding are proposed in \cite{handcraft_win}, which shows that certain \(16 \times 16\) designs outperform Arıkan’s kernel in both exponent and scaling, though at moderately higher complexity. These kernels benefit from structural similarity to the Arıkan matrix, allowing partial reuse of decoding logic while remaining practical for RTPA.

\noindent
\emph{\textbf{Contributions.}} 
We propose \algo, a reinforcement learning framework based on the AlphaZero \cite{alphazero} self-play agent, for the design of polarization kernels with both high error exponent and low decoding complexity under RMLD. 
The design problem is framed as a multi-objective search that jointly targets (i) kernels with a specified partial distance profile and (ii) minimal decoding complexity. 
Our method overcomes the limitations of prior approaches that either rely on exhaustive search \cite{TroBrute24} or restrict the design to fixed kernel structures \cite{TriRTPA23}. 
Numerical simulations show that \algo\, recovers minimal-complexity kernels found via brute-force methods for kernel sizes up to 16. 
This suggests that \algo\, can scale to larger kernels, a regime where traditional search strategies become intractable. 
The proposed approach offers a flexible, data-driven framework for navigating the large and structured space of polarization kernels, and opens the door to automated design of practical polar codes with high performance and low complexity.

\noindent
\textit{Notation:} 
Calligraphic letters denote sets. The set \( \{0, 1, \dots, n-1\} \) is denoted by \( [n] \), and \( \{n, \dots, m\} \triangleq [n-m] \) for \( n < m \), and $\{n, \dots, m-1\} \triangleq [n,m)$.
%
Vectors are written as \( x^n = (x_0, \dots, x_{n-1}) \), and subvectors from index \( k \) to \( n-1 \) are denoted \( x_k^n \). 
Bold lowercase letters (e.g., \( \mathbf{x} \)) represent vectors when the dimension is clear. Random variables are denoted by uppercase letters, and their realizations by lowercase. 
The all-zero matrix of size \( p \times q \) is denoted \( \zerov^{p \times q} \). 


   

  
    




\section{Preliminaries}

\subsection{Polar Codes}
\label{sec:Polar Codes}

A polar code over \( \mathbb{F}_q \) is defined by a transformation \( c_0^{n-1} = \hat{u}_0^{n-1} K^{\otimes m} \), where \( K \) is a non-singular \( \ell \times \ell \) kernel matrix, and \( n = \ell^m \). 
The vector \( \hat{u}_0^{n-1} \in \mathbb{F}_q^n \) includes a frozen set \( \mathcal{F} \subset [n] \) of fixed positions where \( \hat{u}_i = 0 \) for \( i \in \mathcal{F} \), while the remaining positions carry information symbols. 
Polarization conditions for binary kernels were established in \cite{KorErrExp09} and extended to the non-binary case in \cite{mori2014source}. 
Throughout this work, we focus on binary polar codes, i.e., \( q = 2 \), and on the use of identical kernels across all levels of the transformation.
\subsection{Error Exponent for Large Kernels}
\label{sec:Error exponent for large kernel}

Two common criteria for evaluating the performance of a polarization kernel \( K \) are the scaling exponent \cite{HasScaExp2x214, FazScaExp14} and the error exponent \cite{KorErrExp09}. 
In this work, we focus on optimizing kernels with respect to the error exponent.

Let \( W: \{0,1\} \rightarrow \mathcal{Y} \) be a symmetric binary-input discrete memoryless channel (B-DMC) with capacity \( I(W) \).
%
Also 
let $Z_m$ be the Bhattacharyya parameter of a subchannel chosen uniformly at random among those induced by the polar transform \( K^{\otimes m} \). Then, the kernel \( K \) is said to have error exponent \( E(K) \) if:
\begin{itemize}
    \item[(i)] For any \( \beta < E(K) \), \( \lim_{m \to \infty} \Pr[ Z_m \leq 2^{-\ell^m \beta} ] = I(W) \).
    \item[(ii)] For any \( \beta > E(K) \), \( \lim_{m \to \infty} \Pr[ Z_m \geq 2^{-\ell^m \beta} ] = 1 \).
\end{itemize}

It was shown in \cite{KorErrExp09} that, if the frozen set selects the \( n-k \) worst subchannels, then for \( \beta < E(K) \), the block error probability under successive cancellation decoding satisfies \( P_e(\mathcal{C}, W) \leq 2^{-n^\beta} \) for sufficiently large \( n \).

\subsection{Upper Bounds on the Partial Distance Profile (PDP)}
\label{sec:PDP}

In \cite{pdp_bound}, a linear programming (LP) based method is proposed to upper bound the best achievable error exponent for a kernel of size \( \ell \). This upper bound is associated with a partial distance profile (PDP) sequence \( \bar{\Dv}(\ell) = (\bar{D}_0, \dots, \bar{D}_{\ell-1}) \). For any actual kernel \( K \), the corresponding PDP is denoted by \( \Dv(K) = (D_0, \dots, D_{\ell-1}) \), where
\[
D_i = \min_{c \in \langle K_{i+1}^{\ell-1} \rangle} d_H(K_i, c), \quad 0 \leq i < \ell,
\]
with \( K_i \) the \( i \)-th row of \( K \), and \( d_H \) the Hamming distance. The corresponding error exponent is
$E(K) = \frac{1}{\ell} \sum_{i=0}^{\ell-1} \log_\ell D_i.$

While every achievable PDP \( \Dv(K) \) satisfies \( \Dv(K) \leq \bar{\Dv}(\ell) \), the converse does not hold. Thus, kernel search methods often use relaxed profiles \( \widetilde{\Dv}(\ell) \leq \bar{\Dv}(\ell) \) as design targets. The relaxed profiles used in our experiments for \( \ell \in [5,16] \) are reported in Table~\ref{tab:pdp}.

\begin{table}[tbp]
\caption{Relaxed PDP \( \widetilde{\Dv}(\ell) \) and corresponding error exponents in Sec. \ref{sec:Error exponent for large kernel}.}
\begin{center}
\begin{tabular}{|c|l|c|c|}
\hline
\( \ell \) & \rule{0pt}{2.3ex} \( \widetilde{\Dv}(\ell) \) & \( E(\widetilde{\Dv}) \) & \( E(\bar{\Dv}) \) \\
\hline
5  & 1,2,2,2,4 & 0.4307 & 0.4307 \\
6  & 1,2,2,2,4,4 & 0.4513 & 0.4513 \\
7  & 1,2,2,2,4,4,4 & 0.4580 & 0.4580 \\
8  & 1,2,2,2,4,4,4,8 & 0.5000 & 0.5000 \\
9  & 1,2,2,2,2,4,4,6,6 & 0.4616 & 0.4616 \\
10 & 1,2,2,2,2,4,4,4,6,8 & 0.4692 & 0.4692 \\
11 & 1,2,2,2,2,4,4,4,6,6,8 & 0.4775 & 0.4775 \\
12 & 1,2,2,2,2,4,4,4,4,6,6,12 & 0.4825 & 0.4961 \\
13 & 1,2,2,2,2,4,4,4,4,6,6,8,10 & 0.4883 & 0.5005 \\
14 & 1,2,2,2,2,4,4,4,4,6,6,8,8,8 & 0.4910 & 0.5019 \\
15 & 1,2,2,2,2,4,4,4,4,6,6,8,8,8,8 & 0.4978 & 0.5077 \\
16 & 1,2,2,2,2,4,4,4,4,6,6,8,8,8,8,16 & 0.5183 & 0.5274 \\
\hline
\end{tabular}
\label{tab:pdp}
\end{center}
\end{table}

\subsection{Recursive Maximum Likelihood Decoding (RMLD)}
\label{sec:Recursive Maximum Likelihood Decoding}

RMLD is a trellis-based decoding algorithm for binary linear block codes that achieves maximum likelihood (ML) performance with reduced complexity \cite{FujRMLD98}. 
It avoids constructing the full code trellis by recursively building metric tables using small one-section trellises with limited state and branch complexity. 
Two procedures drive the recursion: a base initialization and a recursive update. 
Compared to conventional Viterbi decoding, RMLD achieves significant complexity reduction while maintaining ML optimality.

In \cite{TriRTPA23}, this idea is extended to polar codes with large kernels via the Recursive Trellis Processing Algorithm (RTPA), which exploits the structure of kernel submatrices. 
RTPA aligns with the recursive structure of polar codes, viewing them as generalized concatenated codes 
composed of non-systematic inner codes. 
Likelihood computation is reformulated as soft-input soft-output decoding over an extended trellis, where decoding a bit reduces to identifying the most probable codeword under a final-symbol constraint—efficiently solvable via the Viterbi algorithm.

Next let us describe the RMLD algorithm. 
Due to space limitation, some details and the algorithm pseudocode are omitted. We refer the interest reader to \cite{ForneyMITnotes,FujRMLD98,TriRTPA23}.
For a given polarization kernel $K$ of size $\ell$ by $\ell$, RMLD decoding comprises of $\ell$ decoding phases.  
By reusing the max-tree at different phases, we avoid running the trellis at every phase, thereby reducing the decoding complexity. 
The complexity can be further reduced using the special case optimization technique in \cite{TriRTPA23}. However, for the sake of clarity of presentation, we do not consider these special case techniques in the current implementation. 
Also, for the same reason, we consider the reuse of the trellis corresponding to the largest section of the kernel.
%

A description of the RTPA is provided as follows. 
A given size-$\ell$ kernel $G_\ell$ undergoes $\ell$ decoding phases. 
Each decoding phase corresponds to decoding an extended kernel $G^{(i)}_\ell$, for  $i \in [\ell]$.
The extended kernel $G^{(i)}_\ell$ is constructed by removing the first $i$ rows of $G_\ell$, and appending a column of dimension $(\ell-i, 1)$ to the right. This appended column contains a leading $1$ at index $0$ and zeros elsewhere.   
   
For example, the $F_2$ Arıkan kernel has 2 extended kernels.   
$
G_2^{(0)} = 
\left[
\begin{array}{c c|c}
1&0&1\\
1&1&0\\
\end{array}
\right]
$, and  $
G_2^{(1)} = 
\left[
\begin{array}{c c|c}
1&1&1\\
\end{array}
\right]
$.    
\vspace{0.1cm}
Let $G$ be a generator matrix. Define $G^{p}_{xy}$ as the punctured code of $G$ over the interval  $[x,y)$, and $G^s_{xy}$ as the shortened code over the interval $[x,y)$, meaning that for $G^s_{xy}$, the values outside the interval $[x,y)$ must be zero.  We can divide the interval $[x,y)$ into two subintervals $[x,z)$ and $[z,y)$, where  $x<z<y$. Then $G^p_{xy}$ can be represented as 
\ea{\label{eq:RTPA_matrix}
G^p_{xy} = 
\left[
\begin{array}{*{2}c}
G^{s}_{xz}&0\\
0&G^{s}_{zy}\\
G^{w(0)}_{xy}&G^{w(1)}_{xy}\\
\hline
G^{v(0)}_{xy}&G^{v(1)}_{xy}\\
\end{array}
\right]
=
\left[
\begin{array}{*{1}c}
G^{s}_{xy}\\
\hline
G^{v}_{xy}\\
\end{array}
\right]
}
where $G^v_{xy}= [G^{v(0)}_{xy}, G^{v(1)}_{xy}]$ denotes the subcode of $G^p_{xy}$ that  does not correspond to the shortened code over the section $[x,y)$. $G^{w}_{xy}=[G^{w(0)}_{xy}, G^{w(1)}_{xy}]$ denotes the subcode of $G^s_{xy}$ consisting of codewords that belong to neither the shortened code over the section $[x,z)$ nor that over $[z,y)$.  

We can apply this operation recursively to $G^p_{xz}$ and $G^p_{zy}$. The RMLD algorithm uses a divide-and-conquer approach in its decoding process, where the  decoding output of sections $[x,z)$ and $[z,y)$ are combined to obtain the decoding output for section $[x,y)$.  
%
%
 In phase $i$, an extended kernel $G_\ell^{(i)}$ of size ($\ell-i,\ell+1$) can be punctured into $G^p_{0\ell}$ as shown in  \eqref{eq:RTPA_matrix}. This matrix can then be recursively divided into two sections until the section length becomes one.   That is, 
\ea{
G^p_{xy}\rightarrow [G^p_{xz},G^p_{zy}]
\label{eq:recursive split}
}
where $z=\frac{x+y}{2}$ and the initial values of $x$ and $y$ are $0$ and $\ell$.  
Using $G^p_{xy}, G^p_{xz}$ and $G^p_{zy}$, we can build a binary table -- which we term as the $(w,v,a,b)$ table as in \cite{TriRTPA23}-- that links $G^{v}_{xy}, G^{w}_{xy}$  and $G^{v}_{xz}, G^v_{zy}$.   

\subsection{RMLD Complexity Calculation}
\label{sec:RMLD_complexity}

The proposed kernel design minimizes the decoding complexity of the RMLD algorithm by accounting for its recursive structure. 
As discussed in Sec.~\ref{sec:Recursive Maximum Likelihood Decoding}, RMLD follows a divide-and-conquer strategy: to decode a section \( [x,y) \), it recursively decodes sub-sections \( [x,z) \) and \( [z,y) \), and then combines the results.
The complexity of this combination step depends on the number of rows in the matrices \( G^w_{xy} \) and \( G^v_{xy} \). 
We denote this combination complexity as \( \Csf_{\text{comb}}(G^p_{xy}) \), and express it as:
\begin{equation}
\Csf_{\text{comb}}(G^p_{xy}) = 2^{w+v} + \sum_{k=1}^{w} 2^k,
\label{eq:C comb}
\end{equation}
where \( w \) and \( v \) are the number of rows in \( G^w_{xy} \) and \( G^v_{xy} \), respectively.

The total complexity to decode a section \( [x,y) \) is then the sum of the complexities of the two subsections and the combination step:
\begin{equation}
\Csf(G^p_{xy}) 
= \Csf(G^p_{xz}) + \Csf(G^p_{zy}) + \Csf_{\text{comb}}(G^p_{xy}) 
\label{eq:comp 2}
\end{equation}

The term \( 2^{w+v} \) in \eqref{eq:C comb} accounts for the number of summations required to combine all possible path metrics from the two subsections (i.e., the size of the $(w,v,a,b)$ table \cite{TriRTPA23}), while \( \sum_{k=1}^{w} 2^k \) reflects the number of comparisons required to identify the maximum via a tree search.

The total decoding complexity of a kernel \( G_{\ell} \) is then obtained by summing over all \( \ell \) decoding phases, each corresponding to a different extended kernel \( G^{(i)}_\ell \):
\begin{equation}
\Csf(G_\ell) = \sum_{i=0}^{\ell-1} \Csf(G^{(i)}_\ell) = \sum_{i=0}^{\ell-1} \Csf(G^{(i)p}_{0\ell}).
\label{eq:rmld_tot_comp}
\end{equation}

\paragraph*{Trellis reuse}  
Complexity can be reduced if trellis sections are reused across decoding phases. In particular, if:
\begin{enumerate}
    \item the shortened codes \( G^s_{xz} \) and \( G^s_{zy} \) are identical in phases \( i \) and \( i+1 \), and
    \item the matrices \( G^w_{xy} \) and \( G^v_{xy} \) in phase \( i+1 \) are subcodes of those in phase \( i \),
\end{enumerate}
then the trellis built for phase \( i \) can be reused in phase \( i+1 \), making its computational cost effectively zero.

In principle, one could generalize this idea and reuse arbitrary trellis sections across phases. However, in our implementation we restrict reuse to contiguous phases for two reasons: (i) it simplifies implementation, and (ii) it avoids the need for storing intermediate trellis states. 

Trellis reuse introduces a trade-off between computational and memory complexity: broader reuse reduces computation but increases memory requirements. This breaks the one-to-one coupling between time and space cost present in the baseline version of the algorithm and would require reformulating the optimization objective to handle both resources jointly.

\subsection{Low-complexity RMLD Kernel Search Problem Formulation}
\label{sec:Problem Formulation}

Having introduced polar codes in Sec.~\ref{sec:Polar Codes}, and having specified the design criteria in terms of (i) the PDP, as discussed in Sec.~\ref{sec:Error exponent for large kernel}, and (ii) the decoding complexity under RMLD, as detailed in Sec.~\ref{sec:Recursive Maximum Likelihood Decoding}, we are now ready to formally define our design objective.

Our goal is to design a kernel \( K \) that exhibits low complexity under RMLD decoding while also approaching the PDP upper bound described in Sec.~\ref{sec:PDP}.
Formally, the optimization problem is:
\ea{
K^* & = \argmin_{K \in \mathbb{F}_2^{\ell \times \ell}: \: \Dv(K) \leq  \widetilde{\Dv}(\ell)} \Csf (K).
\label{eq:K minimize C}
}
 The inequality in the constraint is understood element-wise.
We refer to the optimization problem in \eqref{eq:K minimize C} as the \emph{Low-complexity RMLD kernel search problem}.
In the next section, we propose a solution to \eqref{eq:K minimize C} based on a reinforcement learning approach. This choice is motivated by the fact that, for sufficiently large \( \ell \), handcrafted kernel designs—such as those in \cite{TriTre19,tri64}—are no longer effective due to the exponential growth of the search space.

\section{Proposed Approach: \algo}
\label{sec:Proposed Approach}

\subsection{Kerner Search Algorithms}

As a first step, we aim to identify feasible kernels—i.e., kernels whose partial distance profile (PDP) matches the relaxed target \( \widetilde{\Dv}(\ell) \) given in Table~\ref{tab:pdp}. 
While several kernel design methods in the literature focus on maximizing the error exponent, we target both feasibility and complexity. 
Due to the discrete nature of the PDP, however, it is non-trivial to define or measure the “distance” between a candidate kernel's PDP and the ideal bound \( \bar{\Dv}(\ell) \).
We therefore consider two practical search methods: brute-force kernel construction and random agent sampling, each described next.

\subsubsection{Brute-Force Kernel Search \cite{TroBrute24}}
\label{sec:brute_force}

We use the brute-force algorithm of \cite{TroBrute24} to verify whether a kernel exists that satisfies a given relaxed PDP \( \widetilde{\Dv}(\ell) \). 
The kernel is constructed row by row, starting from the bottom row \( i = \ell-1 \) and moving upward.
Let \( \mathcal{M}_w \) denote the set of binary vectors of Hamming weight \( w \), sorted in lexicographic order. 
At row \( i \), the algorithm searches for a vector \( v_i \in \mathcal{M}_{\widetilde{D}_i} \) such that
\[
d(v_i, \langle v_{i+1}, \dots, v_{\ell-1} \rangle) = \widetilde{D}_i,
\]
i.e., the distance between \( v_i \) and the code generated by lower rows matches the PDP requirement. If such a \( v_i \) exists, the algorithm proceeds to row \( i-1 \); otherwise, it backtracks to row \( i+1 \) and selects a new candidate.

This backtracking process ensures completeness: the algorithm either finds a feasible kernel or exhausts all configurations under a predefined step limit.

\subsubsection{Random Agent Kernel Search}
\label{sec:random_agent_search}

In contrast to the deterministic strategy above, we also consider a randomized agent that constructs a kernel by selecting bit positions randomly rather than lexicographically. 
This search is implemented using the same PDP-driven environment described in Sec.~\ref{sec:brute_force}, but replaces systematic enumeration with uniform sampling.

At each step, the random agent chooses a bit in the current row to set to 1. The process continues row by row, checking PDP constraints after each addition. If the full kernel meets \( \widetilde{\Dv}(\ell) \), its RMLD decoding complexity is computed; otherwise, the agent continues until a maximum number of trials is reached.

This method provides a stochastic alternative to brute-force search and is useful for sampling the distribution of decoding complexities among valid kernels. 
In particular, it gives insight into the typical performance of feasible kernels and helps gauge whether low-complexity solutions are rare or common for a given PDP.

\subsection{\algo}\label{sec:polarzero}

The proposed approach to the design of a low-complexity RMLD kernels -- which we term \algo -- relies on AlphaZero  RL agent \cite{alphazero}.

%
 AlphaZero \cite{alphazero} is a self-play RL agent that has mastered various board games without relying on human expertise. 
 The AlphaZero framework consists of two main phases: a self-play phase and a neural network training phase. During the self-play phase, the agent selects actions using the Monte Carlo Tree Search (MCTS) algorithm. The data collected from self-play is then used to train the neural network.
 To accelerate training, we adopt the Gumbel AlphaZero algorithm \cite{gumble_az}, which reduces the search budget required during the self-play phase. 

%

The kernel search environment implemented by \algo \space is described in Algorithm~\ref{alg:AZenv} from a high-level perspective.

Starting from an all-zero $\ell \times \ell$ kernel $G$, the algorithm attempts to construct $G$ row by row while ensuring that the desired PDP $D$ is satisfied.
Within the \texttt{actorThink} function, the RL agent receives the current row-reversed kernel $Gr$ as input and uses the MCTS algorithm to select an action—namely, the index of the column to set to 1.
For each row $Gr[i]$, if its Hamming weight equals the target partial distance $Dr[i]$, the actual partial distance $w$ is computed. If $w = Dr[i]$, the row is accepted, and the agent moves to the next row. Otherwise, the row is reset to zero.
The kernel search environment in Algorithm~\ref{alg:AZenv} returns a \texttt{stateList}, which stores the tuples (state, action, reward) encountered during the episode. This list is then used as training data for the policy and value networks in the AlphaZero framework, as described in Algorithm~\ref{Alg:train_loop}.

To search for low-complexity polarization kernels, we integrate the techniques described in the following subsections into this environment.

\begin{algorithm}
\caption{AlphaZero Self-Play Episode for Kernel Search}\label{alg:AZenv} 
\begin{algorithmic}[1]
\Require Target PDP $D$, neural network $f_\theta$
\Ensure List of (state, action, reward) tuples for training

\State $Dr = \text{reverse}(D)$ \Comment{Row-reversed PDP}
\State $Gr = \zerov^{\ell \times \ell}$ \Comment{Row-reversed kernel}
\State $i = 0$ \Comment{Current row index (top-down)}
\State steps = 0
\State stateList = [ ] \Comment{To store (state, action, reward)}
\While{$i < \ell$ and steps $<$ gameLimit}
    \State steps $\gets$ steps + 1
    \State reward $\gets -c$ \Comment{Step penalty}
    \State $j \gets \text{MCTS}(Gr, f_\theta)$ \Comment{Select column index}
    \State $Gr[i, j] \gets 1$
    
    \If{$\text{weight}(Gr[i]) == Dr[i]$}
        \State $w \gets d(Gr[i], \langle Gr_0^{i-1} \rangle)$
        \If{$w == Dr[i]$}
            \State reward $\gets \alpha$
            \State $i \gets i + 1$ \Comment{Proceed to next row}
        \Else
            \State $Gr[i] \gets 0^\ell$ \Comment{Reset row}
        \EndIf
    \EndIf

    \If{$i == \ell$}
        \State $G \gets \text{reverseRows}(Gr)$
        \State comp $\gets \text{RMLD}(G)$
        \State reward $\gets$ trans(comp, $\gamma$) 
    \EndIf
    \State stateList.append((Gr.copy(), j, reward))
\EndWhile


\State \Return stateList
\end{algorithmic}
\end{algorithm}

\begin{algorithm}
\caption{AlphaZero Training Loop}\label{Alg:train_loop}
\begin{algorithmic}[1]
\For{episode = 1 to $N$}
    \State $\text{episodeData} \gets \text{AlphaZeroKernelSearch}(D, f_\theta)$  \Comment{Run one episode and collect (state, action, reward)}
    \State $\text{trainingData}.append(\text{episodeData})$
    \If{episode mod $K = 0$}
        \State \text{Update  } $f_\theta$ \text{ using accumulated } $\text{trainingData}$
    \EndIf
\EndFor
\end{algorithmic}
\end{algorithm}


\subsection{Randomized initial steps}
To prevent the agent from repeatedly exploring identical kernels, we introduce randomness in the initialization phase by assigning some 1s to selected entries in the lower rows of the kernel $G$. Specifically, in Algorithm~\ref{alg:AZenv}, after setting $Gr = 0^{\ell \times \ell}$, we randomly set a few bits in the top rows of $Gr$ to 1.   
  
For example, consider the case $\ell = 4$ and PDP $D = [1, 2, 2, 4]$ (with reversed version $Dr = [4, 2, 2, 1]$). Since the bottom row $G[3]$ (or its row-reverse version $Gr[0]$) must give $Dr[0]=4$, it is deterministically initialized as $G[3]=[1,1,1,1]$. The search then starts from the second-to-last row $G[2]$, corresponding to $Dr[1]=2$. To introduce diversity, we assign $G[2,j]=1$ (or its row-reverse version $Gr[1,j] = 1$) where $j$ is chosen at random. For instance, if $j = 1$, the initial kernel could be:
$$
Gr = 
\left[\begin{array}{cccc}
1 & 1 & 1 & 1 \\
0 & 1 & 0 & 0 \\
0 & 0 & 0 & 0 \\
0 & 0 & 0 & 0 \\
\end{array}\right] 
\begin{array}{c}
  \\
\quad \leftarrow \text{current row } i=1  \\
  \\
  \\
\end{array}
$$

\subsection{Structure of the Rewards}   

The RL algorithm uses the following reward structure:
\begin{itemize}[leftmargin=*]
  \item Negative step reward ($c$): To encourage faster convergence, we penalize the RL agent for each step taken by assigning a negative reward of $-c$ per step, with $c > 0$. However, $c$ must be chosen carefully: if it is too large, the agent may prioritize minimizing the number of steps over reducing kernel complexity. In our experiments, we set $c = 0.1$.
  \item New row reward ($\alpha$) : Every time the RL agent finds a new row that satisfies the PDP constraint, it receives a reward $\alpha > 0$. If the agent successfully constructs all $\ell$ rows, the total reward from this term is $\ell \cdot \alpha$. We set $\alpha = 5$ for $\ell = 12$ and $\alpha = 10$ for $\ell = 16$.   
  \item Complexity reward: The reward based on kernel decoding complexity is computed using the transformation function: 
\begin{equation} \label{eq:comp_trans}
\begin{split}
  \text{trans}(\text{comp},\gamma) = R_{\min} + (R_{\max} - R_{\min})  \\
  \cdot\left(\frac{\text{comp}_{\max} - \text{comp}}{\text{comp}_{\max}-\text{comp}_{\min}} \right)^{\gamma} 
\end{split}
\end{equation}
where $R_{\min}$ and $R_{\max}$ denote the minimum and maximum reward, and $\text{comp}_{\min}$ and $\text{comp}_{\max}$ represent the observed bounds on kernel complexity. The value of $\text{comp}_{\max}$ is estimated using the random agent from Sec. \ref{sec:random_agent_search}. For instance, for $\ell=16$, we set $\text{comp}_{\max}=5000$, based on a maximum observed complexity of $5690$. The value of $\text{comp}_{\min}$ is determined from the lowest complexity kernel found so far; for $\ell=16$, we set $\text{comp}_{\min} = 1300$.   We set $R_{\min}=0$ and $R_{\max}=\text{comp}_{\max} - \text{comp}_{\min}$. 
The parameter $\gamma$ controls the nonlinearity of the reward: higher values emphasize low-complexity kernels. We use $\gamma = 2$.   
\end{itemize}

The total reward $v$ obtained upon reaching the top row of a kernel that satisfies the target PDP $D$ is given by:  
\begin{equation} \label{eq:reward_shaping}
 v =- c \cdot (\text{steps}-\ell) + \text{trans}(\text{comp},\gamma) + \ell\cdot \alpha.
 \end{equation}

\begin{figure}
    \centering
    \resizebox{0.48\textwidth}{!}{\input{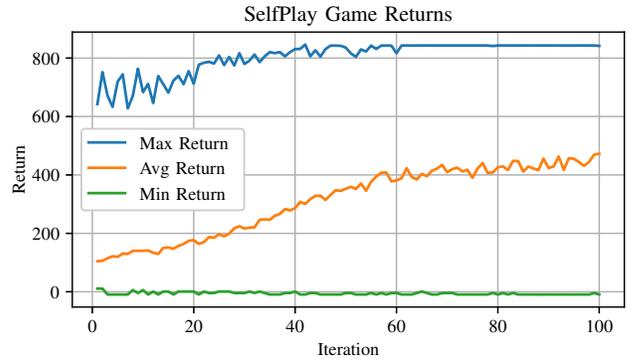}}
    \vspace{-0.35cm}
    \caption{Minimum, maximum, and average total rewards during training for $\ell=12$. Each iteration consists of $2000$ self-play games, with a total of  $2\cdot 10^5$ episodes over $100$ iterations.}
    \label{fig_12}
    \vspace{-0.1cm}
\end{figure}

\section{Numerical Results}

\subsection{Kernel Search Methods}
In this section, we present numerical results on the decoding complexity of polarization kernels discovered using two methods: the random agent introduced in Sec.~\ref{sec:random_agent_search}, and the AlphaZero-based algorithm \algo{} described in Sec.~\ref{sec:polarzero}.

\subsection{Random Agent}

Table~\ref{tab:rand_comp} reports the minimum and maximum decoding complexities (measured via RMLD) of kernels of size $\ell$ obtained using the random agent search. For each value of $\ell$, we also indicate the number of Monte Carlo (MC) iterations performed.

As expected, the maximum complexity increases rapidly with kernel size $\ell$, approximately following an exponential trend. This behavior highlights the growing difficulty of discovering low-complexity kernels through unguided random exploration.
\begin{table}[tbp]
\caption{Minimum and maximum RMLD complexity found by the random agent in Sec. \ref{sec:random_agent_search}.}
\begin{center}
\begin{tabular}{|c|c|c|c||c|c|c|c|}
\hline
\textbf{$\ell$} & Min & Max & Iterations & \textbf{$\ell$} & Min & Max & Iterations \\ 
\hline
4  & 32   & 40   & 10k   & 11 & 523  & 1041 & 10k \\
5  & 51   & 93   & 10k   & 12 & 668  & 1448 & 200k \\
6  & 84   & 146  & 10k   & 13 & 1049 & 2109 & 10k \\
7  & 117  & 219  & 10k   & 14 & 1424 & 2834 & 10k \\
8  & 152  & 280  & 10k  & 15 & 1717 & 4107 & 10k \\
9  & 271  & 503  & 10k   & 16 & 1774 & 5690 & 400k \\
10 & 326  & 708  & 10k   &    &      &      &     \\
\hline
\end{tabular}
\label{tab:rand_comp}
\end{center}
\vspace{-0.75cm}
\end{table}
\subsection{\algo{} Agent }
We now focus on the results obtained\footnote{The codebase used for the experiments in this section is publicly available to support reproducibility: \texttt{https://github.com/jaco267/AlphaPolar}.
} using the \algo{} agent for two kernel sizes: $\ell = 12$ and $\ell = 16$.

\begin{figure}
    \centering
    \resizebox{0.48\textwidth}{!}{\input{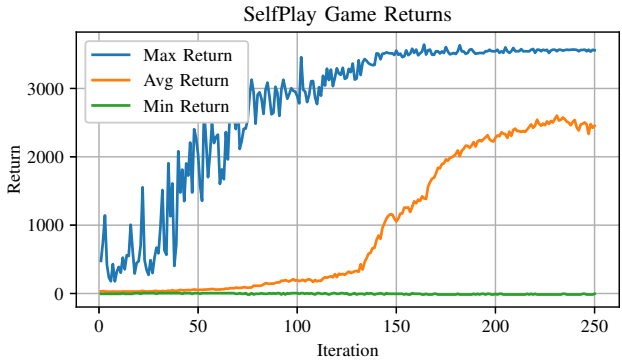}}
    \vspace{-0.5cm}
    \caption{Minimum, maximum, and average total rewards during training for $\ell=16$. Each iteration consists of 2000 self-play games, with a total of $5 \cdot 10^5$ episodes over 250 iterations.}
    \label{fig_16}
    \vspace{-0.5cm}
\end{figure}
Fig.~\ref{fig_16} shows the evolution of total rewards during training for $\ell = 16$ and with PDP  $\widetilde{\Dv}(16)$ from Table~\ref{tab:pdp}. At each iteration, $K = 2000$ self-play games are generated (see Algorithm~\ref{Alg:train_loop}), and the neural network $f_\theta$ is trained. The average reward stabilizes after approximately 230 iterations, suggesting convergence. For smaller values of $\ell$, convergence occurs faster; for instance, for $\ell = 12$, convergence is reached after around 100 iterations.

The best kernel found for $\ell = 16$ has decoding complexity 1396, comparable to the handcrafted kernel from \cite{handcraft_win}, which achieves complexity 1384 under our RMLD complexity evaluation (see Sec.~\ref{sec:RMLD_complexity}). The \algo-found kernel is:
\begin{center}
\scalebox{0.7}{$
A_{16} =
\left[
\begin{array}{*{16}c}
1&0&0&0&0&0&0&0&0&0&0&0&0&0&0&0\\
0&0&1&0&0&0&0&0&0&0&0&1&0&0&0&0\\
0&0&0&1&0&1&0&0&0&0&0&0&0&0&0&0\\
0&0&0&0&1&0&1&0&0&0&0&0&0&0&0&0\\
0&0&0&0&1&1&0&0&0&0&0&0&0&0&0&0\\
0&1&0&0&1&0&0&0&1&0&0&0&0&0&1&0\\
1&0&0&1&0&1&0&1&0&0&0&0&0&0&0&0\\
1&1&0&0&0&0&1&1&0&0&0&0&0&0&0&0\\
1&1&1&1&0&0&0&0&0&0&0&0&0&0&0&0\\
0&1&0&1&1&1&0&0&1&0&0&0&1&0&0&0\\
0&0&0&0&1&0&0&1&1&0&0&0&1&1&1&0\\
1&1&1&1&1&1&1&1&0&0&0&0&0&0&0&0\\
0&0&1&1&0&0&1&1&1&1&1&1&0&0&0&0\\
1&1&1&1&0&0&0&0&0&0&1&1&0&1&1&0\\
1&0&0&1&1&0&1&0&0&1&1&0&0&1&0&1\\
1&1&1&1&1&1&1&1&1&1&1&1&1&1&1&1\\
\end{array}
\right]
$}
\end{center}
This kernel was discovered starting from the initial state:
\begin{center}
\scalebox{0.75}{$
A_{16}[14:15] = 
\left[
\begin{array}{*{16}c}
0&0&0&1&1&0&0&0&0&1&1&0&0&0&0&0\\
1&1&1&1&1&1&1&1&1&1&1&1&1&1&1&1\\
\end{array}
\right]
$}
\end{center}
Similarly, for $\ell=12$, the best decoding complexity observed is 652. The rows of the corresponding  kernel $A_{12}$ are reported in the footnote\footnote{$A_{12}=[$\texttt{0x800};
\texttt{0x210};
\texttt{0xA00};
\texttt{0x240};
\texttt{0x003};
\texttt{0xA50};
\texttt{0x162};
\texttt{0x868};
\texttt{0x464};
\texttt{0xE4B};
\texttt{0x78C};
\texttt{0xFFF}$]$} in hexadecimal notation.

We now compare the BLER performance of $(n=256, k=128)$ codes constructed using: (i) Arıkan’s $\ell=2$ kernel; (ii) the handcrafted kernel $G_{16}$ from \cite{handcraft_win}; and (iii) the \algo-found kernel $A_{16}$. The frozen bit sets are optimized at each SNR. Arıkan’s code is decoded using SCD, while $G_{16}$ and $A_{16}$ are decoded using RMLD.

\begin{figure}
    \centering
    \begin{tikzpicture}
\begin{axis}[
    xlabel={$E_b/N_0$ [dB]},
    ylabel={BLER},
    ymode=log,
    xmin=0, xmax=4.5,
    ymin=1e-5, ymax=1,
    legend pos=south west,
    grid=both,
    width=8cm,
    height=6cm
]

\addplot+[mark=*] coordinates {
    (0.0, 8.4202e-01)
    (0.5, 6.6234e-01)
    (1.0, 4.1924e-01)
    (1.5, 2.1348e-01)
    (2.0, 7.8460e-02)
    (2.5, 2.1560e-02)
    (3.0, 4.0400e-03)
    (3.5, 4.6000e-04)
    (4.0, 7.0000e-05)
    (4.5, 2.0000e-05)
};
\addlegendentry{RMLD $G_{16}$}

\addplot+[mark=triangle*] coordinates {
    (0.0, 8.2158e-01)
    (0.5, 6.2598e-01)
    (1.0, 3.8660e-01)
    (1.5, 1.8264e-01)
    (2.0, 6.1300e-02)
    (2.5, 1.5260e-02)
    (3.0, 2.8400e-03)
    (3.5, 4.5000e-04)
    (4.0, 2.0000e-05)
    (4.5, 5.0000e-05)
};
\addlegendentry{RMLD $A_{16}$}

\addplot+[mark=square*] coordinates {
    (0.0, 8.8312e-01)
    (0.5, 7.3074e-01)
    (1.0, 5.1536e-01)
    (1.5, 2.9604e-01)
    (2.0, 1.2870e-01)
    (2.5, 4.5820e-02)
    (3.0, 1.2370e-02)
    (3.5, 2.3000e-03)
    (4.0, 4.0000e-04)
    (4.5, 9.0000e-05)
};
\addlegendentry{Arıkan's $2\times 2$ kernel}

\end{axis}
\end{tikzpicture}
    \vspace{-0.25cm}
    \caption{Block Error Rate (BLER) performance of polar codes using Arıkan’s kernel (SC decoding), the handcrafted kernel $G_{16}$ from \cite{handcraft_win}, and the \algo-found kernel $A_{16}$. At each SNR, frozen bits are selected based on $10^5$ MC simulations. BLER curves are estimated using $10^5$ MC iterations per SNR.}
    \label{fig_perf_A16}
    \vspace{-0.5cm}
\end{figure}
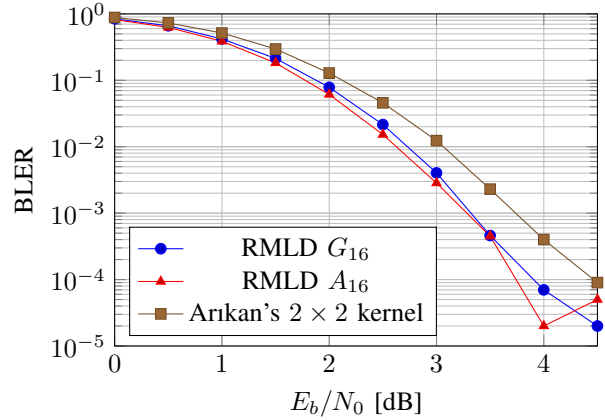

As shown in Fig.~\ref{fig_perf_A16}, the kernels $G_{16}$ and $A_{16}$ achieve similar BLER performance, both outperforming Arıkan’s code. This improvement is consistent with their superior error exponents, $E_{G_{16}} = E_{A_{16}} \approx 0.5183$, compared to $E_{\text{Arıkan}} = 0.5$.

\bibliographystyle{ieeetr}
\bibliography{main_bib}

\end{document}